\documentclass[twocolumn,trackchanges]{aastex631}

\submitjournal{ApJ}

\shortauthors{Takahashi et al.}

\begin{document}

\newcommand{\lya}{${\rm Ly\alpha}$}
\newcommand{\ha}{${\rm H\alpha}$}
\newcommand{\hb}{${\rm H\beta}$}
\newcommand{\hg}{${\rm H\gamma}$}
\newcommand{\hd}{${\rm H\delta}$}

\def\msun{{\rm M}$_{\odot}$ }
\def\um{$\mu${\rm m}}
\def\kms{{\rm km/s}}
\newcommand{\oi}{[\textrm{O}~\textsc{i}]}
\newcommand{\oii}{[\textrm{O}~\textsc{ii}]} 
\newcommand{\oiii}{[\textrm{O}~\textsc{iii}]}
\newcommand{\nii}{[\textrm{N}~\textsc{ii}]}
\newcommand{\sii}{[\textrm{S}~\textsc{ii}]}
\newcommand{\heii}{[\textrm{He}~\textsc{ii}]}
\newcommand{\nieii}{[\textrm{Ne}~\textsc{iii}]}
\newcommand{\cii}{[\textrm{C}~\textsc{ii}]}
\renewcommand{\ttdefault}{pcr} 
\def\bfr{\bf \color{red}}

\title{Green Flash: Residual Emissions Enshrouded in Low-mass Balmer-break Galaxies at $z\sim5$}

\correspondingauthor{Kosuke Takahashi}
\email{kosuke.takahashi@astr.tohoku.ac.jp}
\author[0009-0009-8116-0316]{Kosuke Takahashi}
\affiliation{Astronomical Institute, Tohoku University, 6-3, Aramaki, Aoba, Sendai, Miyagi 980-8578, Japan}

\author[0000-0002-8512-1404]{Takahiro Morishita}
\affiliation{IPAC, California Institute of Technology, MC 314-6, 1200 E. California Boulevard, Pasadena, CA 91125, USA}

\author[0000-0002-2993-1576]{Tadayuki Kodama}
\affiliation{Astronomical Institute, Tohoku University, 6-3, Aramaki, Aoba, Sendai, Miyagi 980-8578, Japan}

\author[0009-0002-8965-1303]{Zhaoran Liu}
\affiliation{Astronomical Institute, Tohoku University, 6-3, Aramaki, Aoba, Sendai, Miyagi 980-8578, Japan}

\author[0000-0002-9509-2774]{Kazuki Daikuhara}
\affiliation{Astronomical Institute, Tohoku University, 6-3, Aramaki, Aoba, Sendai, Miyagi 980-8578, Japan}
\affiliation{Institute of Space and Astronautical Science, Japan Aerospace Exploration Agency, 3-1-1, Yoshinodai, Chuou-ku, Sagamihara, Kanagawa 252-5210, Japan}

\author[0000-0002-0486-5242]{Nuo Chen}
\affiliation{Astronomical Institute, Tohoku University, 6-3, Aramaki, Aoba, Sendai, Miyagi 980-8578, Japan}




\begin{abstract}
Recent James-Webb Space Telescope (JWST) observations have discovered galaxies that are already passively evolving at $z>4$, $\sim1.5$\,Gyr after the Big Bang. Remarkably, some of these galaxies exhibit strong emission lines such as \ha\ and \oiii\ while showing a strong continuum break at $\sim3650$\,\AA\ i.e., Balmer break, giving us a unique insight into the physical mechanisms responsible for early galaxy quenching. In this study, we investigate the nature of four such galaxies at $z=5.10$--$5.78$ identified in the Abell~2744 field, using JWST/NIRCam and NIRSpec data. Our spectral energy distribution (SED) fitting analysis reveals that these galaxies have been mostly quiescent since $\sim100$\,Myr prior to the observed time. We find a higher dust attenuation in the nebular component than in the continuum in all cases. This suggests the presence of dusty star-forming regions or obscured AGN, which could be a {\it residual} signature of past quenching. For one of the galaxies with sufficient medium-band coverage, we derive the \hb+\oiii\ emission line map, finding that the line-emitting region is located in the center and is more compact ($R_e=0.7$\,kpc) than the stellar component ($R_e=0.9$\,kpc). For this specific galaxy, we discuss a scenario where quenching proceeds in the manner of ``outside-in", a stark contrast to the inside-out quenching commonly seen in massive galaxies at later cosmic times.
\end{abstract}



\section{Introduction} \label{sec:intro}
Revealing the mechanisms behind galaxy quenching is crucial for understanding galaxy evolution across cosmic time. Recent studies have advanced our knowledge of quenching processes, highlighting the roles of both environmental factors and internal mechanisms.

One significant factor of the internal effects is active galactic nuclei (AGN) feedback. The AGNs, powered by gas accretion onto the central supermassive black holes, can inject vast amounts of energy into their host galaxies. This energy can heat the surrounding gas and/or drive powerful outflows, expelling gas from the galaxies and effectively halting star formation. This process, known as the AGN-driven quenching or ``negative feedback”, has been supported by theoretical studies \citep[e.g.,][]{hopkins09, Dubois2013}.

Observations have confirmed massive galaxies already quenched their star formation relatively early in the Universe's history \citep[e.g.][]{glazebrook_massive_2017, schreiber_near_2018, tanaka_stellar_2019, valentino_quiescent_2020, ito_x-ray-detected_2024}. \cite{glazebrook_massive_2017} reported the discovery of a massive ($>10^{11}\,M_\odot$) quiescent galaxy at a redshift of $z = 3.717$, corresponding to just 1.6 billion years after the Big Bang. It suggests that rapid quenching mechanisms, possibly involving intense AGN feedback or starburst-driven winds, were already in place in the early Universe. 

Moreover, recent studies with JWST have found massive quiescent galaxies at $z \sim 5$ \citep[e.g.,][]{de_graaff_efficient_2024, carnall_massive_2023, kakimoto_massive_2024}, and above \citep{Weibel25}. These galaxies often exhibit a broad \ha\ emission line \citep{carnall_massive_2023} or a strong \oiii$_{\lambda\lambda4959,5007}$ line \citep{de_graaff_efficient_2024}, suggesting the presence of AGNs at the center, and that AGN feedback mainly suppresses their star-forming activities.

The low-mass regime of high-$z$ galaxies, on the other hand, remained relatively unexplored until recently. Low-mass galaxies are susceptible also to supernova feedback, which heats or removes the gas and suppresses star formation. \cite{strait_extremely_2023} discovered a low-mass galaxy $(M_{\ast}\sim10^{7.6}~\mathrm{M_{\odot}})$ at $z = 5.2$ that has features of Balmer-break and \ha\ + \nii\ lines, but there was no conclusive evidence to support the presence of AGN. The Balmer break, which is a characteristic feature of stellar populations dominated by A-type stars, is an indicator of recently quenched star formation i.e. $\sim0.1–1.0$ Gyr prior to the observed epoch. Even further at $z=7.3$, a low-mass Balmer-break galaxy named as ``JADES-GS-z7-01-QU" $(M_{\ast}\sim10^{8.7}~\mathrm{M_{\odot}})$ that shows very weak emission lines was reported in \cite{looser_recently_2024}.

Recent theoretical studies have investigated the quenching mechanisms of low-mass Balmer-break galaxies in the early Universe \citep[e.g.,][]{lovell_first_2023, dome_mini-quenching_2023, gelli_quiescent_2023, faisst_dead_2024}. 
\citet{gelli_quiescent_2023} constructed low-mass  ($\lesssim10^{9.5}~\mathrm{M_\odot}$) galaxy samples using the SERRA simulation and found that rapid quenching driven by radiative winds from massive stars and/or AGNs is required to reproduce the spectra of JADES-GS-z7-01-QU. 
In addition to rapid quenching, bursty and short-timescale star formation (lasting $\sim40$ Myr) can also occur at $z = 4-8$ \citep{dome_mini-quenching_2023}. 
Despite these implications, the SED of JADES-GS-z7-01-QU could not be fully reproduced by cosmological simulations, such as SERRA \citep{gelli_quiescent_2023}, IllustrisTNG, and VELA \citep{dome_mini-quenching_2023}.
\citet{faisst_dead_2024} showed that incorporating both stochastic star-formation histories and dust obscuration at star-forming regions can reproduce the SED of JADES-GS-z7-01-QU. 
However, a definitive conclusion has not yet been reached, especially the role of AGN activity. 
Further observational evidence is required to fully understand the quenching mechanisms at high redshift in the low-mass regime.

\begin{figure*}
    \includegraphics[width=1.0\textwidth]{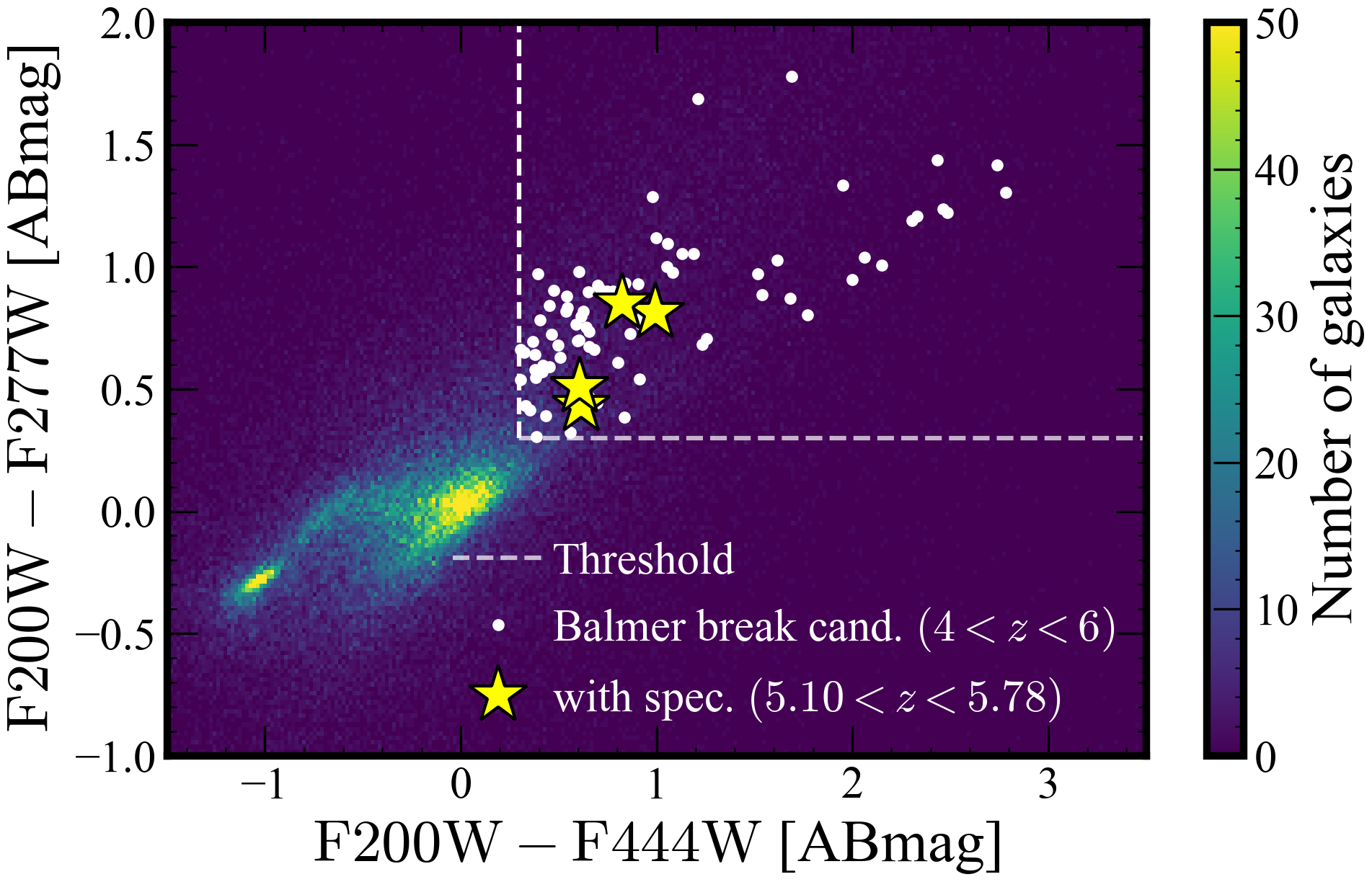}
    \includegraphics[width=1.00\textwidth]{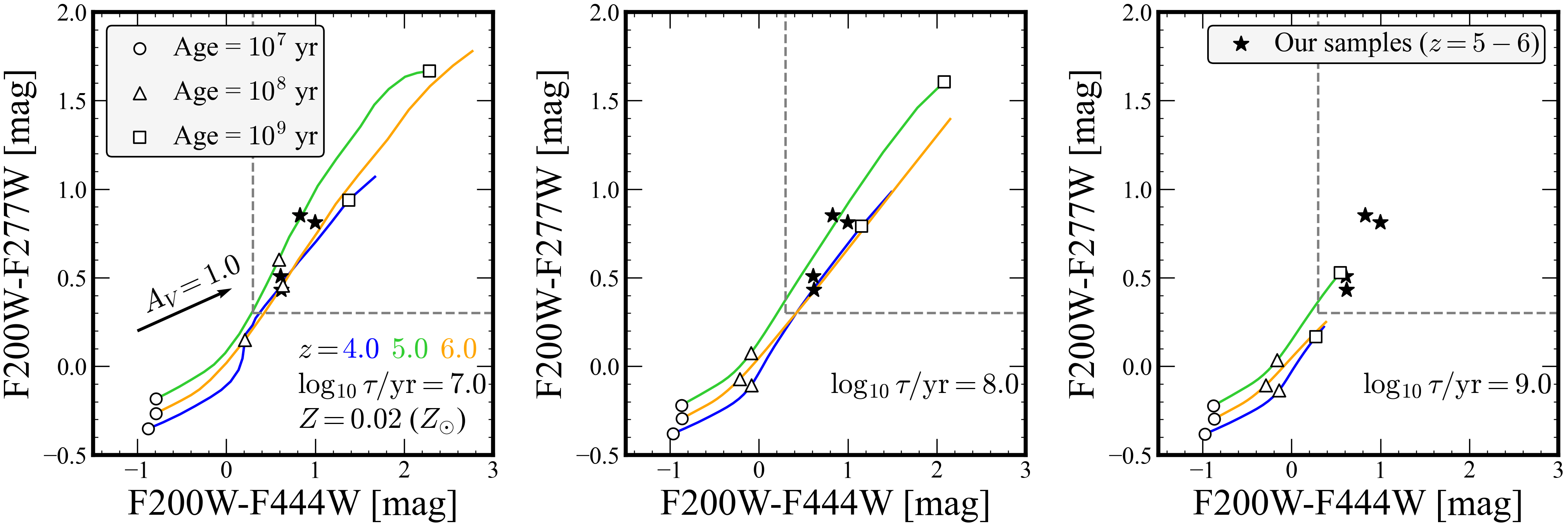}
    \caption{(Top): color-color (F200W-F444W and F200W-F277W) diagram. The density map shows the all objects in the parent catalog. White dots are the candidates of Balmer break galaxies at $4 < z_\mathrm{phot} < 6$. Yellow stars indicate four spectroscopically confirmed Balmer break galaxies at $5.10\leq z_\mathrm{spec}\leq5.78$ (this study).
    (Bottom): simulated color evolutionary tracks, generated by \texttt{CIGALE} code for the delayed-$\tau$ star formation history and the solar metallicity. Each panel sets different $\tau$, the length of star formation. 
    White circles, triangles, and squares indicate the stellar age of $10^7$yr, $10^8$yr and $10^9$yr, respectively. 
    The black star markers represent our four samples that contain spectral data.}
    \label{fig:colormodel}
\end{figure*}

While research on quiescent galaxies at high redshift is advancing, one key factor remains largely unexplored: the exact spatial location of the latest star formation within the galaxy before quenching. With the latest high-resolution observations, this question has become increasingly accessible. This is crucial to our understanding of quenching models, which may involve a wide range of spatial scales, from black hole accretion disks to the galaxy-scale physical mechanisms. The question of how and where star formation stops remains entirely open. 

In this regard, this study focuses on galaxies that have recently ceased star formation, to gain insight into the mechanisms responsible for early quenching in the Universe. In particular, we identify four Balmer-break galaxies at $z\sim5$ from recent JWST surveys and study their properties, aiming to uncover quenching mechanisms at the low-mass end, $\lesssim 10^9\,M_\odot$. 
In addition to existing deep imaging and spectroscopic data from JWST, gravitational lensing by the foreground cluster Abell2744 enables us to gain a more comprehensive picture of quenching across different masses.

This paper is structured as follows:
In Section \ref{sec:methods}, we describe the data set we use and the details of the analyses we perform.
In Section \ref{sec:results}, we show the results from the SED fitting and emission line analyses.
In Section \ref{sec:discussions}, we discuss the location of quenching within a galaxy and possible mechanisms.
In Section \ref{sec:conclusions}, we summarize the nature of these galaxies.

Throughout this paper, we use the AB magnitude system \citep{oke_secondary_1983}, and assume the flat $\Lambda$CDM model with $H_0=70$ $\mathrm{km~s^{-1}~Mpc^{-3}}$, $\Omega_{m,0}=0.3$ and $\Omega_{\Lambda,0}=0.7$, and the Chabrier initial mass function (IMF) \citep{chabrier_galactic_2003}.

\section{Methods}
\label{sec:methods}
\subsection{Imaging and Spectroscopic Data}
We utilize JWST dataset available in the Abell 2744 field, which is a known lensing cluster at $z = 0.308$. For our study, we use publicly available images produced by the GLASS-JWST team, originally designed to investigate extremely high redshift($z\geq10$) lensed galaxies \citep{merlin_early_2022,paris_glass-jwst_2023}. The dataset includes images from 8 broad-band filters and 12 medium-band filters taken by NIRCam (F070W, F090W, F115W, F140M, F150W, F162M, F182M, F200W, F210M, F250M, F277W, F300M, F335M, F356W, F360M, F410M, F430M, F444W, F460M, F480M). We also combine HST photometry (ACS: F435W, F606W, F775W, F814W; WFC3-IR: F105W, F125W, F140W, F160W) taken in multiple HST programs \citep{postman_cluster_2012, lotz_frontier_2017, kelly_extreme_2018, steinhardt_buffalo_2020}. Details of the imaging data reduction are described in \cite{morishita_accelerated_2025}. 

We also use the spectroscopic data taken by JWST/NIRSpec Micro Shutter Array (MSA) using the PRISM/CLEAR configuration ($\mathrm{R} \sim 30-300$; wavelength range: $0.6\text{--}5.3~\mu\mathrm{m}$). These data were collected from two observation programs, GO2561 (UNCOVER:PI I. Labbe) and GO3073 \citep{castellano_jwst_2024}, coherently reduced and compiled in \citet{roberts-borsani_between_2024}. 

Because the observation targets cannot be perfectly centered within their spectral slits, moderate flux is lost outside the slit because of the small MSA shutter aperture size. Additionally, since the Full Width at Half Maximum (FWHM) of the Point Spread Function (PSF) increases with wavelength, the slit throughput loss also depends on wavelength. 
We scale the spectrum using the photometry to consider the slit loss. 
The absolute flux calibration is obtained by scaling the spectra to match the average flux derived from the F277W, F356W, and F444W photometry.
These galaxies are located at the lensing cluster region. We adopt the magnification factor $\mu$ derived from \cite{bergamini_new_2023, bergamini_glass-jwst_2023} \citep[See also][]{morishita_accelerated_2025}, except for ID85007; since ID85007 is beyond the coverage of the available lens model, we assume no magnification, $\mu=1.0$.

\subsection{Target selection}
\label{selection}
To study Balmer break galaxies at $4 <  z < 6$, we select candidates using the criteria outlined below:

\begin{equation}
\label{eq_balmer1}
    \mathrm{F200W}-\mathrm{F277W}\geq0.30~[\mathrm{mag}]
\end{equation}

\begin{equation}
\label{eq_balmer2}
    \mathrm{F200W}-\mathrm{F444W}\geq0.30~[\mathrm{mag}]
\end{equation}

\begin{equation}
\label{eq_lyman}
\mathrm{S/N}_\mathrm{F115W}\geq2.0~\mathrm{and}~\mathrm{S/N}_\mathrm{F606W}<2.0
\end{equation}

\begin{equation}
\label{eq_detection}
    \mathrm{F444W}\leq27.0~[\mathrm{mag}]~\mathrm{and}~\mathrm{S/N}_\mathrm{F200W}\geq2.0
\end{equation}

Equations \ref{eq_balmer1} and \ref{eq_balmer2} capture the Balmer break at $4 < z < 6$. 
Figure~\ref{fig:colormodel} shows the selection. The red color excess between F200W and F277W can be caused by strong emission lines such as \ha\ at $z \sim 3.2$ and \oiii\ at $z \sim 4.5$, which enter the F277W band. 
A false Balmer break feature could be mimicked due to them. Therefore, we add a further constraint, Equation~\ref{eq_balmer2}, to robustly capture the Balmer break feature. 
Dusty galaxies at different redshift have steep redder slope. But heavily dusty galaxies (e.g. \textit{HST-dark} galaxies) will be removed by Equation \ref{eq_lyman}.   
To limit the brightness of our candidates, we apply Equation \ref{eq_detection}, with the first condition representing the detection threshold and the second ensuring that we can analyze the bluer (shallower) side of the break.
Furthermore, to ensure the selection of candidates at $z > 4$, we apply criteria based on the Lyman break feature (Equation \ref{eq_lyman}). 
There are $\sim 95,000$ objects in the original photometric catalog. 837 objects are selected by these color criteria. 

\begin{figure*}
    \centering
    \includegraphics[width=\textwidth]{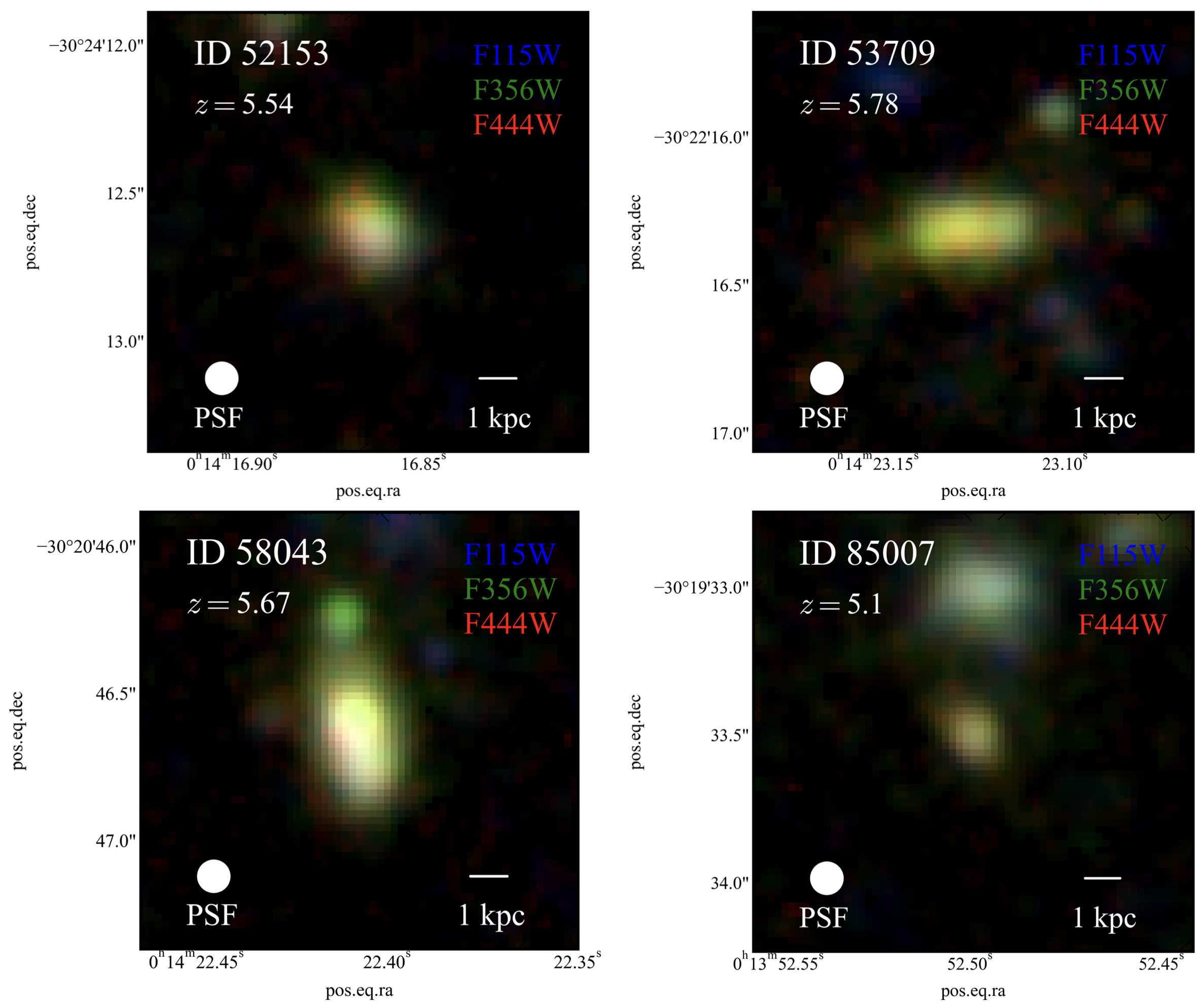}
    \caption{RGB images of our four Balmer break samples. The F115W and F356W images are PSF-matched to the F444W image.}
    \label{fig:thumbnails}
\end{figure*}

In order to discuss the nature of secure Balmer-break galaxies, we cross-match our Balmer break candidates with spectroscopic datasets from programs GO2561 and GO3073, compiled in \citet{roberts-borsani_between_2024}.
Four Balmer-break galaxies are identified, located at $z_\mathrm{spec} = 5.10 - 5.78$. In what follows, we focus on these spectroscopically confirmed galaxies as our main sample in this study, whereas photometric candidates will be further investigated in a future work.
Figure~\ref{fig:thumbnails} shows their RGB images using F115W (blue), F356W (green) and F444W (red).

\subsection{SED fitting analysis}  
\label{SED_fitting}
We conduct SED fitting using \texttt{Bagpipes} \citep{carnall_inferring_2018, carnall_vandels_2019} with photometric and spectroscopic data. We use \texttt{Bagpipes} code to estimate physical properties of galaxies.
We assume the non-parametric SFH \citep{leja_how_2019}. 
\texttt{Bagpipes} uses the Kroupa IMF \citep{kroupa_variation_2001}. 
The stellar population model is \texttt{BC03} \citep{bruzual_stellar_2003}.
\texttt{CLOUDY} \citep{ferland_2017_2017} is also used to include the effect of emission lines. The dust attenuation model is \citep{calzetti_dust_2000}. The dust emission model is the model of \citet{draine_infrared_2007}. We set the parameters mass of formed stars $\log_{10}(M_{\ast}/M_{\odot})\in[5, 12]$, metallicity $\log_{10}(Z/Z_{\odot})\in[0.01, 2.5]$, the ionized parameter $\log_{10}U\in[-3.0, -0.5]$ and dust attenuation $A_V\in[0.0, 4.0]$.

In this paper, we adopt the dust attenuation law in \citet{calzetti_dust_2000} which is $R'_V=4.05\pm0.80$ and $E(B-V)_\mathrm{stellar} = f\times E(B-V)_\mathrm{gas}$ (where $f=0.44\pm0.03$).
The dust attenuation curve $k'(\lambda)$ is described as $k'(\lambda)=A(\lambda)/E(B-V)_\mathrm{stellar}$, where $\lambda$ represents wavelength in the rest frame.
Following \citet{calzetti_dust_2000}, for $0.63~\mu \mathrm{m}\leq\lambda\leq2.20~\mu \mathrm{m}$ (which covers \ha\ and \nii\ lines), we use 
\begin{equation}
      k'(\lambda)=2.659(-1.857+1.040/\lambda)+R'_V
\end{equation}
and for $0.12~\mu \mathrm{m}\leq\lambda\leq0.63~\mu \mathrm{m}$ (which covers \oii, \hb\ and \oiii\ lines),
\begin{equation}
k'(\lambda)=2.659(-2.156+1.509/\lambda-0.198/\lambda^2+0.011/\lambda^3)+R'_V  
\end{equation}
Then, once we know $A_V$ value, we can calculate the intrinsic flux $F_\mathrm{int}$ from the observed flux $F_\mathrm{obs}$ using equation \ref{attenuation}. 
\begin{equation}
\label{attenuation}
F_\mathrm{int}(\lambda)=F_\mathrm{obs}(\lambda)\times10^{0.4k'(\lambda)A_V/R'_V}
\end{equation}

\subsection{Emission line fitting}
\label{sec:method_line_fitting}
We perform the emission line fitting with a single Gaussian using \texttt{emcee} \citep{foreman-mackey_emcee_2013}. 
To estimate the stellar continuum, we utilize model spectra derived from \texttt{Bagpipes}. 
We obtain model spectra of the stellar continuum and create the continuum-subtracted spectra of these galaxies. 
Then we measure the emission line properties (e.g., line flux, luminosity, line width, and $\mathrm{EW_0}$) of the \oii$\lambda\lambda 3726, 3729$, \hb, \oiii$\lambda\lambda4959, 5007$, and \ha+\nii\ lines. 
For the \oiii\ and \ha+\nii\ lines, we also conduct 2- and 3-component Gaussian fitting. Even in low-resolution spectra, multiple-components fittings are useful to constrain the emission line properties (see below).
However, when the Signal-to-Noise ratios are below $\sim2$, the line fitting does not converge. 
Consequently, for those with emission lines undetected at $ < 2\,\sigma$, we estimate 2-$\sigma$ upper limits using the rms estimate in the continuum near the emission line.
In this paper, we assume the Case B situation ($T = 10^4~\mathrm{K}$ and $n_e = 10^{2}~\mathrm{cm^{-3}}$) and use the \ha/\hb\ ratio derived by \cite{osterbrock_astrophysics_2006} (\ha/\hb\ $=2.863$).
We use the PRISM low-resolution ($R \sim 200$ at $\sim 4$ \um) spectroscopic data. For \ha\ emission line at $z = 5.5$, the spectral resolution is $ \sim 1500$ \kms\ and the \ha\ line and the 
\nii\ doublets are partially blended, which potentially affect our interest, the \ha-to-\nii\ line ratio. 
In Sec.~\ref{sec:discussions}, we investigate the limitation of spectral resolution by modeling synthetic spectra assuming different line ratios.
By assuming that the line widths of both \ha\ and \nii\ lines are 230 \kms\ and a \nii\ doublet ratio of \nii$\lambda6583$/\nii$\lambda6548 = 2.96$ \citep{osterbrock_astrophysics_2006}, we find a $2$-$\sigma$ upper limit of \ha\,/\,\nii\,$ < 3$. The measured line fluxes are reported in Table~\ref{line_fit}.

\begin{figure*}
    \centering
    \includegraphics[width=\textwidth]{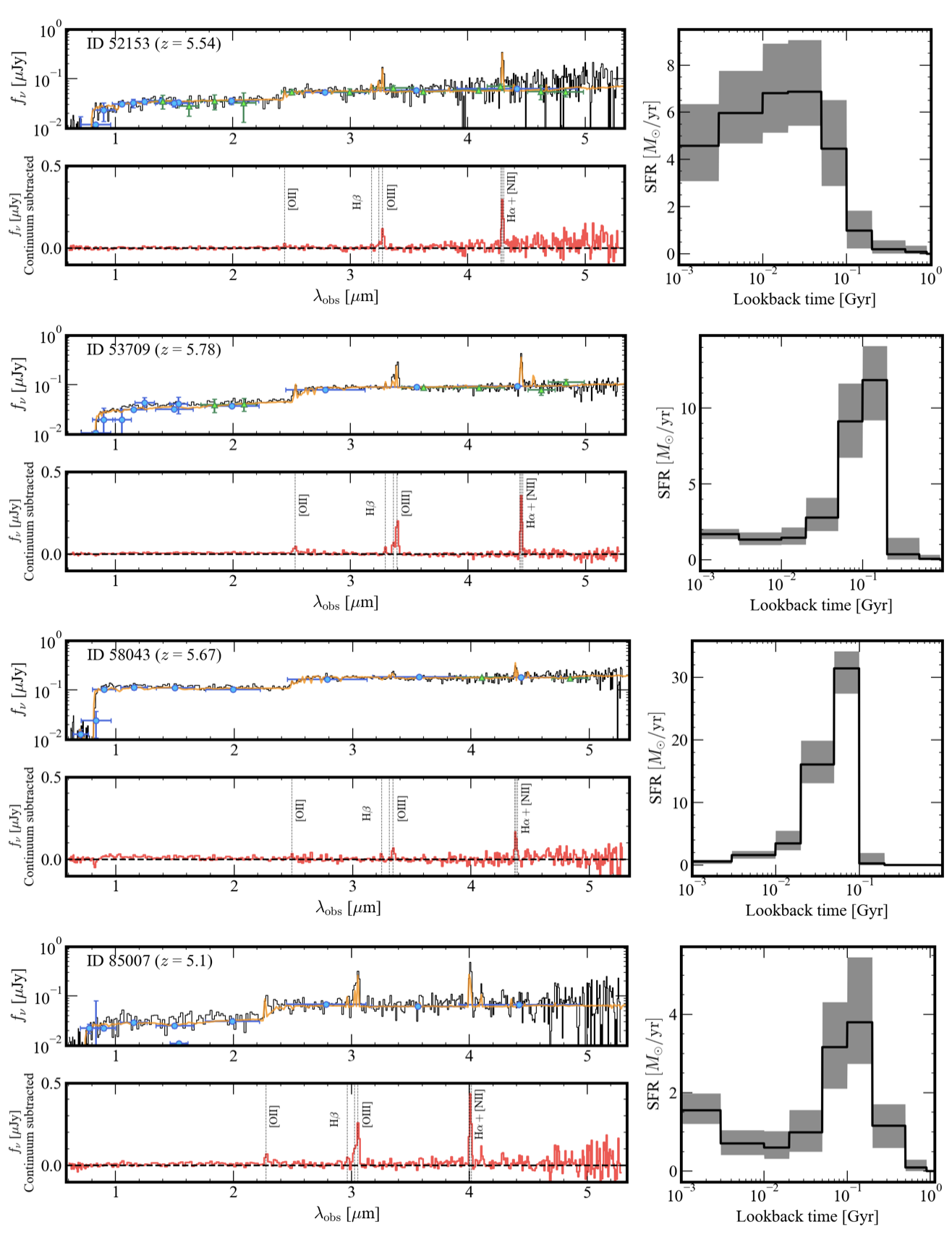}
    \caption{JWST/NIRSpec PRISM spectra of our sample galaxies. 
    Each top panel shows the photometry (blue dots: broad-band filters of HST and JWST, green dots: medium-band filters of JWST) and spectrum (black solid line) of the galaxy. The orange solid line indicates the estimated total spectrum from \texttt{Bagpipes} SED fitting. Fluxes are corrected by magnification.
    Each bottom show the continuum subtracted spectrum, where the continuum is the one modeled by \texttt{Bagpipes}.}
    \label{fig:observation}
\end{figure*}

\section{Results}
\label{sec:results}

\subsection{Stellar mass and star-formation rate}

\label{sec:SED_results}

The SED fitting results derived from \texttt{Bagpipes} are shown in Table \ref{value} and Figure \ref{fig:observation}. 
\texttt{Bagpipes} can treat not only photometry data but also spectrum, allowing for a more reliable derivation of the star formation history.  
The stellar masses of Balmer break galaxies derived from \texttt{Bagpipes} is $M_{\ast,\mathrm{Bagpipes}}=10^{8.8-9.5}~\mathrm{M_{\odot}}$. 
The star formation rates (SFR) are $\mathrm{SFR}_\mathrm{Bagpipes}=1-30~\mathrm{M_{\odot}/yr}$.  
We compare these $M_{\ast}$ and SFR with star formation main sequence (hereafter main sequence) at this redshift ($z=5.5$) shown in Figure \ref{sfms}. 
We use the main sequence derived by \citet{popesso_main_2023} using star-forming galaxies at $z \sim 5.5$. We summarize the results of SED fitting in Table \ref{value}.
This indicates that this galaxy has experienced an increase of star-formation activity once above $z = 6$, then SFR started dropping off $\sim10~\mathrm{Myr}$ ago (See star formation history in each galaxy; Figure \ref{fig:observation}). 
The mass weighted ages derived from \texttt{Bagpipes} SED fitting are consistent with the color track examined by \texttt{CIGALE} delayed-$\tau$ SFH models shown in Figure \ref{fig:colormodel}.

\subsection{Dust attenuation}
We compare the dust attenuation measured from the spectroscopic Balmer decrement (i.e. \ha/\hb) with the one derived from \texttt{Bagpipes} SED fitting ($A_V$).
To illustrate the differences between these two attenuation measurements, we plot the dust-uncorrected \ha/\hb\ ratio against the $A_V$ values obtained from the SED fitting. Except for ID53709, the \hb\ line is not detected in three of the four objects. For this case, we provide a lower limit for the \ha/\hb\ ratio.
For comparison, we also present two empirical relations between the dust-uncorrected \ha/\hb\ ratio and the $A_V$ values using the equation below:
\begin{equation}
\label{ha_hb_ratio_inverse}
(\mathrm{H\alpha}/\mathrm{H\beta})_\mathrm{obs}=(\mathrm{H\alpha}/\mathrm{H\beta})_\mathrm{int}\times10^{-0.4\frac{k'(\mathrm{H\alpha})-k'(\mathrm{H\beta})}{f\times R'_V}  A_V}.
\end{equation}

\begin{figure*}
    \centering
    \includegraphics[width=0.9\textwidth]{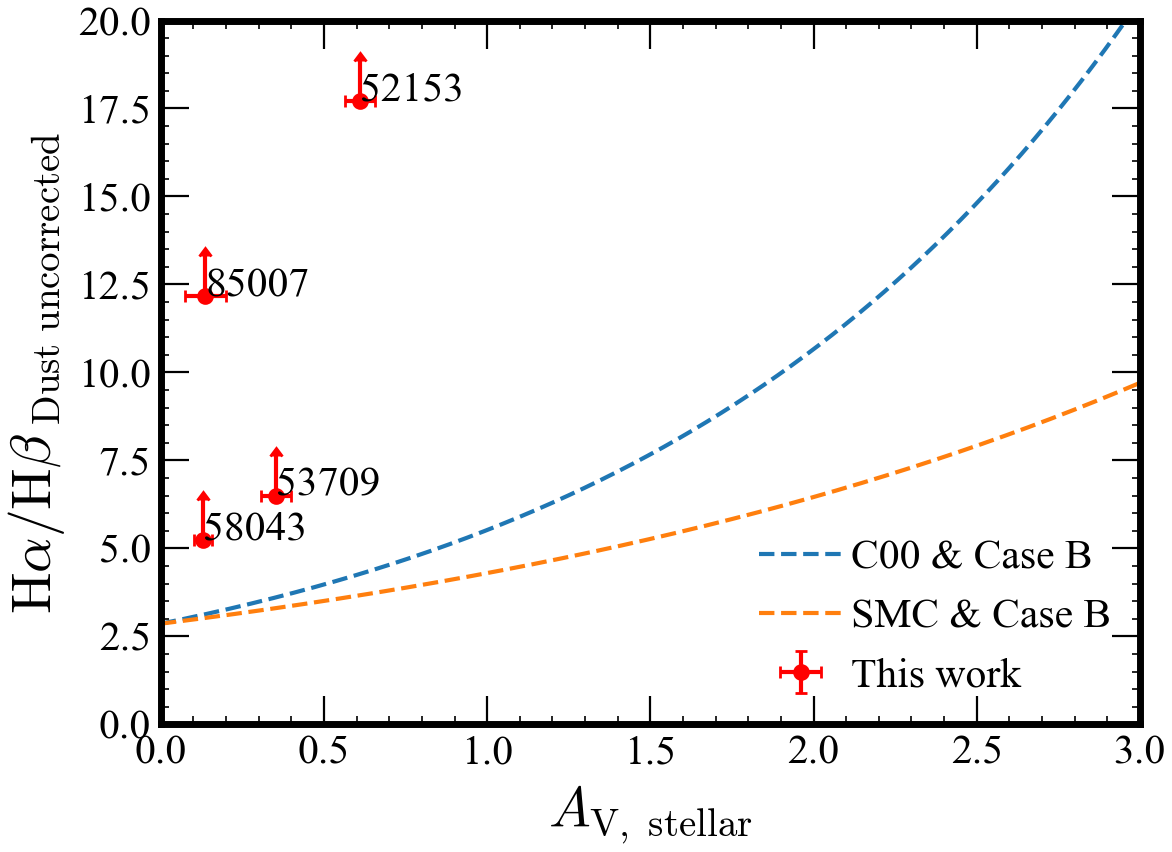}
    \caption{Red dots represent our four Balmer break galaxies. The horizontal axis indicates stellar attenuation $A_V$, derived from SED fitting, while the vertical axis shows the line flux ratio \ha-to-\hb\ (equivalent to nebular extinction $A_\mathrm{V, nebular}$), which has not been corrected for dust attenuation. Note that all \ha-to-\hb\ measurements are lower limits, due to non detection of \hb\ (IDs 52153, 58043 and 85007) and the \ha+\nii\ blending (resulting in a conservative lower flux limit for \ha; Sec.~\ref{sec:method_line_fitting}). The two dashed lines correspond to empirical models: the blue line assumes the Calzetti attenuation law (C00, \citealt{calzetti_dust_2000}) and Case B recombination, while the orange line assumes the attenuation law of the Small Magellanic Cloud (SMC, \citealt{gordon_quantitative_2003}) and Case B recombination. 
    }
    \label{fig:dust_comp}
\end{figure*}

For one of them, we use the combination of the Calzetti dust extinction law \citep{calzetti_dust_2000} and Case-B, as indicated in blue dashed line in Figure \ref{fig:dust_comp}. The other assumes the Small Magellanic Cloud (SMC) dust attenuation law \citep{gordon_quantitative_2003} and Case-B, shown in orange dashed line in the same plot. When a data point is positioned above the empirical line, it would indicate that the attenuation value estimated from the emission lines exceeds that derived from SED fitting, and vice versa. In Figure \ref{fig:dust_comp}, we see that all the four galaxies are located above both empirical lines. 
Given that the MSA slit was placed near the galaxy center in all cases, the finding here suggests that their central region, where most line emissions supposedly originate, is more attenuated than the galaxy as a whole is. We discuss this in further detail in Sec.~\ref{sec:linemap}.

\subsection{The possibility as AGN}
\label{sec:The_possibility_of_AGN}

We first examine the X-ray data taken by Chandra/ACIS. \citealt{bogdan_evidence_2023-1} (Program ID 25926) observed the Abell2744 field to investigate high-redshift AGN, with a total exposure time of 61 ks. Three of our Balmer-break galaxy samples are covered in the obtained image. However, none of them are detected in the image. By using the tools of Chandra data analysis, we measure the full-band ({0.5--7.0\,keV}) photometry. The aperture size is set to 1 arcsec, larger than the resolution of Chandra, 0.5 arcsec. To estimate the background flux, we set the annulus within $r_\mathrm{in}=3.0$ arcsec to $r_\mathrm{in}=6.0$ arcsec. The upper limit flux is $\log_{10}(F_\mathrm{0.5-7~keV}/~\mathrm{[erg/s/cm^2]})<-15.0$, which roughly corresponds to the luminosity $\log_{10}(L_\mathrm{0.5-7~keV}/~\mathrm{[erg/s]})<44.2$ ($1\sigma$ upper limit).
The limit is comparable to the typical X-ray luminosity of AGN at similar redshifts ($\log_{10}(L_\mathrm{0.5-2~keV}/~\mathrm{[erg/s]})\sim44$) \citep[e.g.,][]{pouliasis_active_2024}.

\begin{figure}
    \centering
    \includegraphics[width=0.48\textwidth]{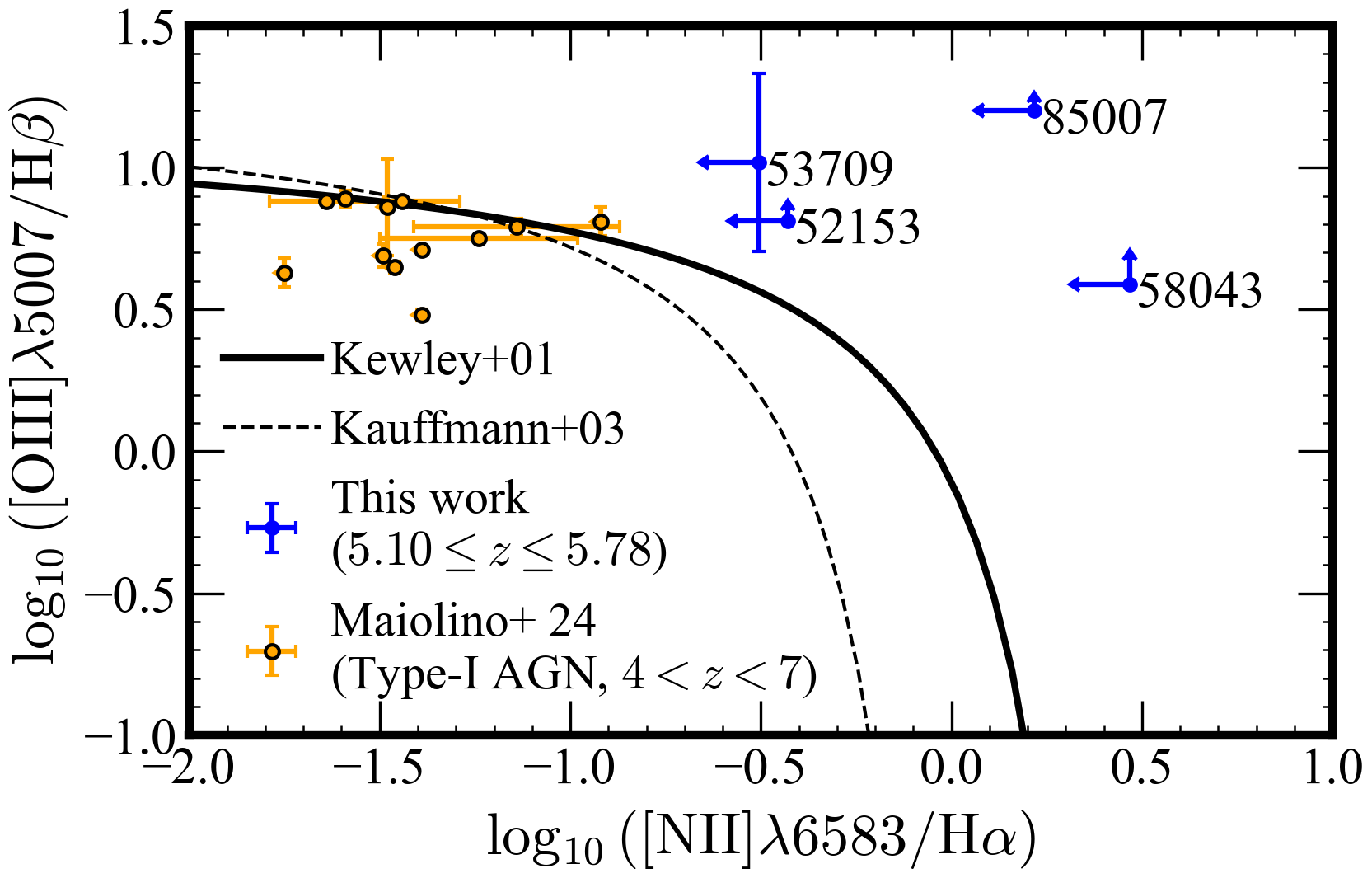}
    \caption{BPT diagram: The blue dots are a lower limit of \oiii/\hb\ ratio and an upper limit of \nii/\ha\ ratio in each galaxy, with arrows indicating the $2\sigma$ upper or lower limits. The black solid and dashed lines are the corresponding separating AGN and star forming galaxies in local Universe from \cite{kewley_theoretical_2001, kauffmann_host_2003}. Orange dots are the \ha\ broad line (Type-I) AGN at $4<z<7$ found by \cite{maiolino_jades_2024-1}.}
    \label{bpt}
\end{figure}

We then inspect the presence of AGNs in the traditional Baldwin, Philips, \& Terlevich diagram \citep[BPT diagram;][]{Baldwin1981}. The flux ratios of \oiii/\hb\ and \nii/\ha\ are measured through the line fitting analysis in Sec.~\ref{sec:method_line_fitting}.
Line ratio measurements are shown in Figure \ref{bpt}. For comparison, we also include the boundaries between star-forming galaxies and AGN in the local Universe, as well as high-$z$ ($4 < z < 7$) AGNs identified as Type-I AGN by \citet{maiolino_jades_2024-1}. At high redshift, galaxies tend to exhibit higher ionization parameters, and most galaxies hosting HII regions show a high \oiii/\hb\ ratio. Consequently, distinguishing star-forming galaxies from AGN-hosting galaxies becomes challenging for high-$z$ samples.

The inferred lower/upper limits (2\,$\sigma$) of our samples are too weak to definitively constrain the presence of an AGN. Together with the absence of significant X-ray detection and the apparent absence of strong \nii\ emissions (see also Sec.~4.1), our analysis excludes the presence of Type-I AGN. In addition, we note that none of the four galaxies is a point source (Fig.~\ref{fig:thumbnails}), which would indicate the presence of a Type-I AGN. Therefore, the observed emission lines are likely attributed to star formation, or Type-II (i.e. dust-obscured) AGN if any.

\begin{figure}
    \centering
    \includegraphics[width=0.5\textwidth]{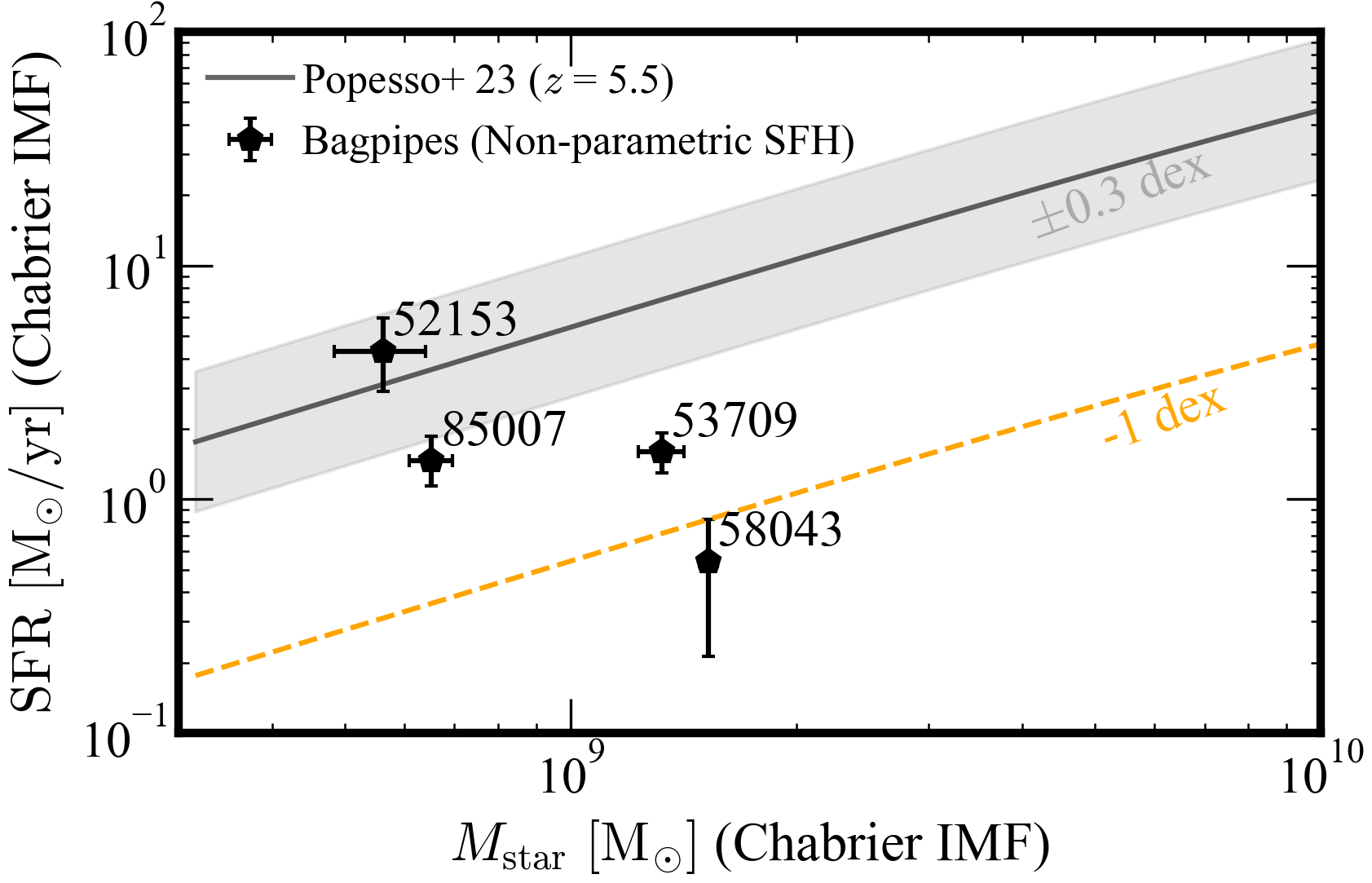}
    \caption{Comparison with main sequence at $z = 5.5$: The black solid line is the star formation main sequence derived from \citep{popesso_main_2023}. Gray shaded region and orange dashed line show the $\pm0.3$ dex and -1 dex from the main sequence, respectively. A black dot is derived from \texttt{Bagpipes}. ID52153 is on the main sequence, but this galaxy's spectrum shows a Balmer break feature, indicating the absence of O- and B-type stars in its whole component.}
    \label{sfms}
\end{figure}

\begin{figure*}
    \centering
    \includegraphics[width=\textwidth]{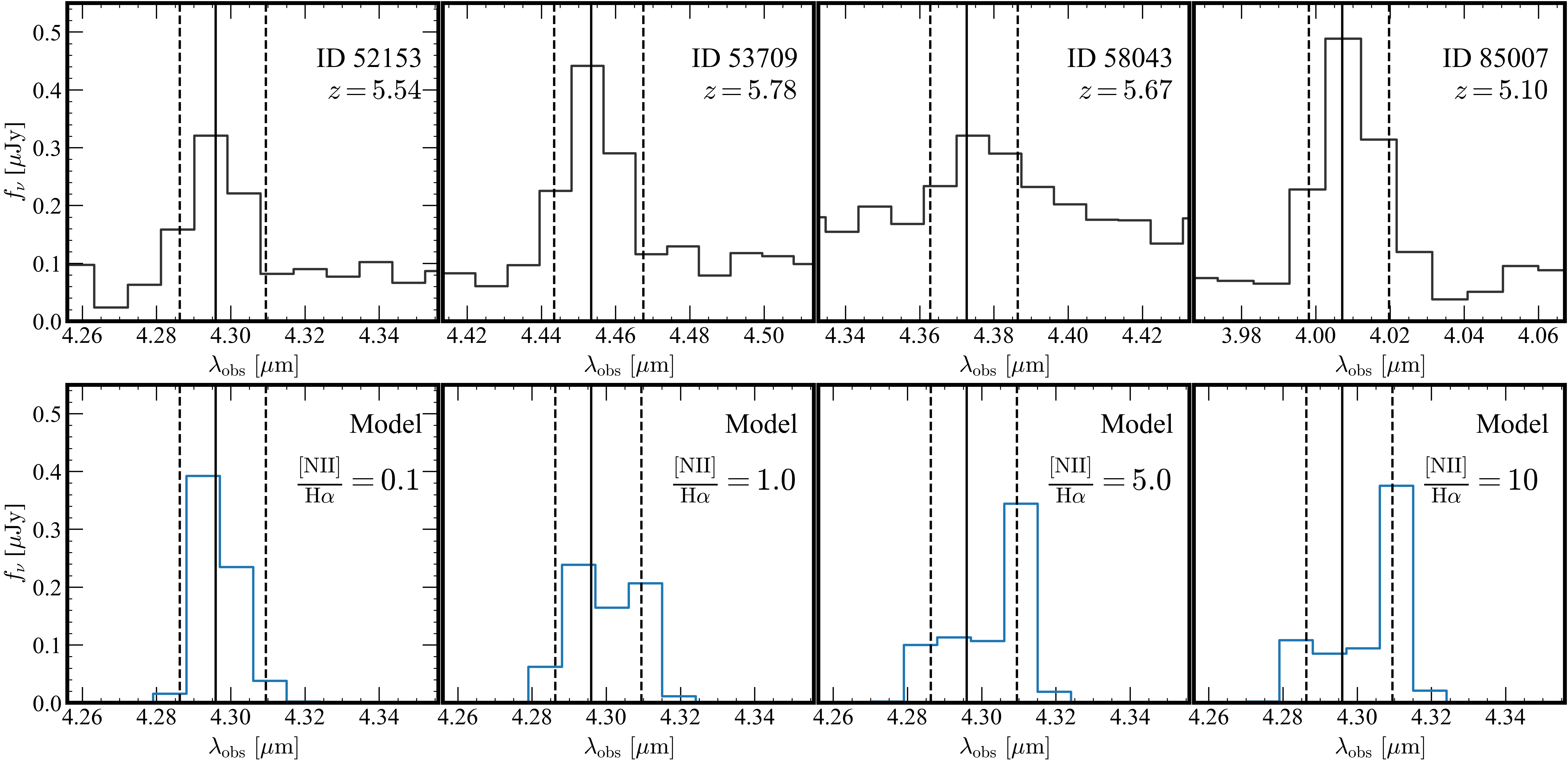}
    \caption{($Top$): Panels show observed spectra of our samples around the \ha+\nii\ emission line complex, partially blended in our PRISM spectra.
    The solid lines are the center of \ha, and the dashed lines show the \nii\ doublet at each redshift. ($Bottom$): Panels demonstrate how the emission lines appear at PRISM resolution ($R\sim200$) as an example of ID52153 when varying the line ratio of \nii\ and \ha. The line width is fixed at 230 km/s, and the \ha\ and \nii\ doublet are each modeled as single Gaussian profiles with three components in total. We fix the total flux ($F_\mathrm{H\alpha+[NII]}$) as $1\times10^{-18}~\mathrm{[erg/s/cm^{2}]}$ in these models. These synthetic line profiles are shown after being degraded to a resolution of about $R\sim200$. }
    \label{fig:ha_nii}
\end{figure*}

\begin{figure*}
    \centering
    \includegraphics[width=\textwidth]{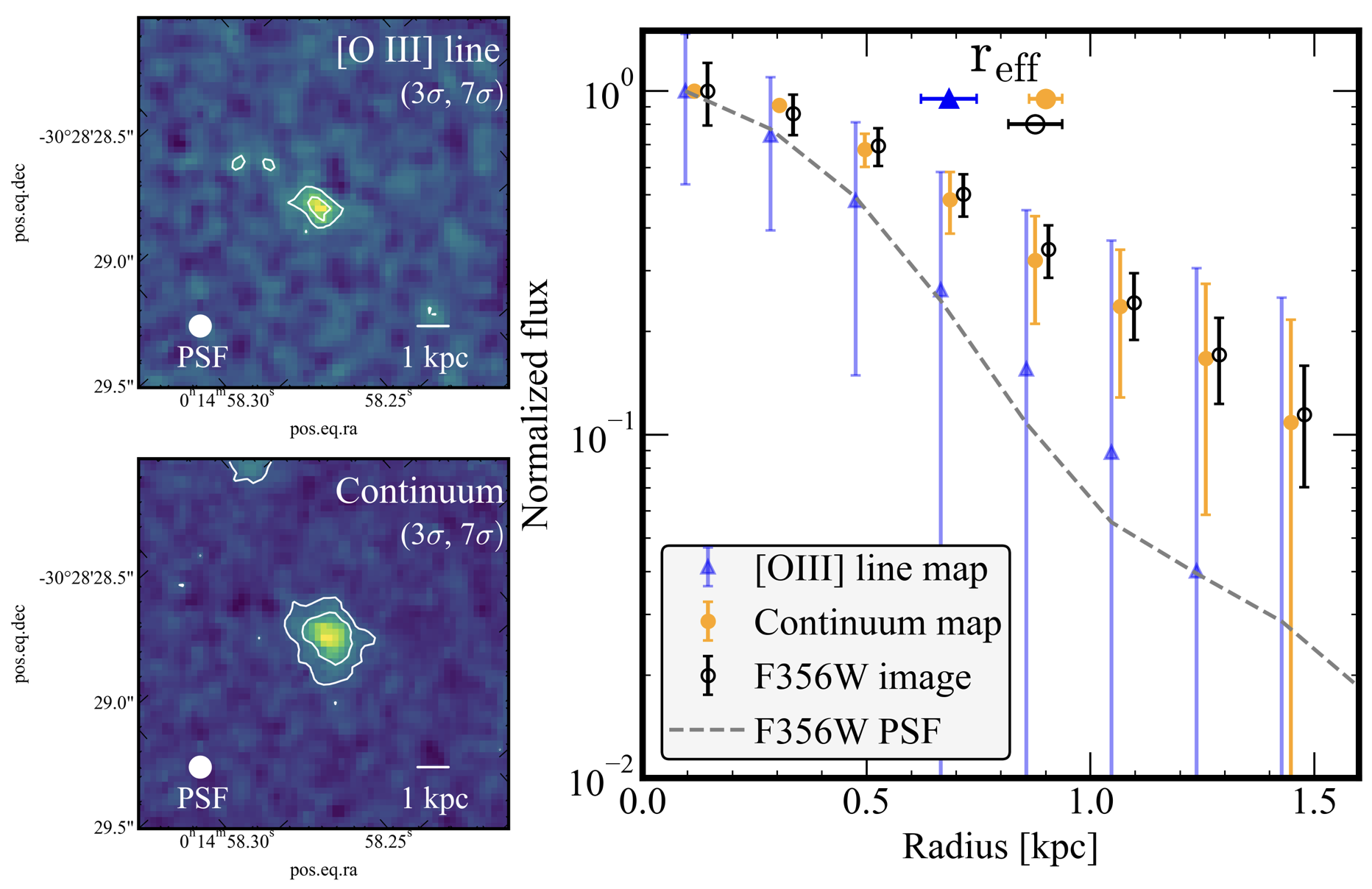}
    \caption{Centrally concentrated \oiii\ emission line. The $2'\times2'$ cutout images of the \oiii\ emission line map (\textit{top left}) and the stellar continuum map (\textit{bottom left}). These maps are constructed from the PSF-matched F335M and F356W images using Equations \ref{lineeq} and \ref{conteq}. The continuum map shows the stellar continuum at the rest-frame $\sim5000~\mathrm{\AA}$. The white contours in each map indicate significance levels ($3\sigma$ and $7\sigma$). The PSF size shown in the bottom-left corner of each map is the empirical PSF size of F356W ($\lambda $$\sim$$3.5~\mathrm{\mu m}$). (\textit{Right}): The radial profiles are shown for the \oiii\ emission line map, the stellar continuum map and the F356W image. The gray dashed line indicates a radial profile of PSF of F356W image. The dots at the top of this figure show the effective radii of each image. The continuum profile (orange dots) is spatially more extended to outer regions than \oiii\ line. 
    }
    \label{fig:jwst_oiii}
\end{figure*}

\section{Discussions}
\label{sec:discussions}

\subsection{Different dust attenuation in nebular and stellar components}
\label{The_difference_of_dust_attenuation}

All four galaxies exhibit a strong \ha\ line ($F_\mathrm{H\alpha}>10^{-18}~\mathrm{[erg/s/cm^{2}]}$), despite the non-detection of the \hb\ line. 

Assuming Case~B and a coherent attenuation across the galaxy, an $A_\mathrm{V, nebula} \gtrsim 2$\,mag would be required to explain the flux difference between \ha\ and \hb\ for all the cases. However, this is clearly inconsistent with the $A_V$ of the stellar component i.e. the one derived from our SED fitting ($\lesssim0.6$\,mag).

The difference in attenuation of the line and stellar components suggests that those components represent distinct physical properties.
We propose two scenarios to explain this;
($i$) the \ha+\nii\ line complex is dominated by the \nii\ doublets, where our \ha\ line flux would be overestimated; ($ii$) Strong \ha\ and \hb\ but higher dust attenuation for nebular emission lines than for stellar light.

To examine Case~($i$), we investigate whether the observed shape of the line complex can be better fitted with stronger \nii. In this case, we adopt the dust attenuation $A_\mathrm{V}$ derived from \texttt{Bagpipes} SED fitting i.e. sets the same as for the stellar attenuation. The \nii\ doublet ratio of \nii$\lambda6583$/\nii$\lambda6548$ = 2.96 and a line width of 230 km/s for \nii$\lambda\lambda6548, 6583$ and \ha\ are also assumed. Using the upper limit of the \hb\ flux, we derive the upper limit of the \ha\ flux. From this, we estimate the \nii/\ha\ ratio using the combined \ha+\nii\ flux. The \nii/\ha\ ratios in our samples are estimated to be 1.82 (ID52153), 0.49 (ID53709), 1.65 (ID58043) and 4.24 (ID85007), with \ha\ comprising less than 35\% (ID52153), 67\% (ID53709), 38\% (ID58043), and 19\% (ID85007) of the each total \ha+\nii\ flux. However, as shown in Figure \ref{fig:ha_nii}, these \nii/\ha\ line ratios cannot reproduce the shapes of the observed spectra of our samples. The qualitative comparison with the synthetic spectra in the same figure suggests that the line ratio is $\sim0.1$. 

Therefore, we conclude that Case~($i$) is unlikely, leaving Case~($ii$) as a more plausible explanation for our samples. We note that in Case~($ii$) a uniform dust distribution across the galaxy is not required. We further address Case~($ii$) using one of our samples in Sec.~\ref{sec:linemap}.

\subsection{Line map and size measurement}\label{sec:linemap}

At first glance, the fact that these four galaxies exhibit both the Balmer-break and strong emission lines appears contradictory. However, we have so far treated these systems as a single-zone system. 
With the arrival of high-resolution deep photometric and spectroscopic observations from JWST/NIRCam, it is now possible to spatially resolve emission lines in distant galaxies \citep[e.g.,][]{GimenezArteaga2023, Matharu24, Liu24, Chen2025}, allowing us to examine spatial distributions of emission-line and stellar continuum.

To further investigate it within our data, we focus on one particular object, ID52153. Unlike the other three, the redshift of this object facilitates the \oiii\,4959+5007 doublet lines to fall within one of the medium-band filters, F335M. While \hb\ also locates within the same filter, since its strength is much smaller than the \oiii\ doublets, we treat it negligible here.
Combined with the broadband filter F356W, whose wavelength coverage encloses that of F335M, it allows us to create both an emission-line map and a continuum map as follows:

\begin{equation}
\label{lineeq}
    F_\mathrm{line} = (f_\mathrm{MB} - f_\mathrm{BB})\frac{\Delta_\mathrm{MB}}{1-\Delta_\mathrm{MB} / \Delta_\mathrm{BB}}
\end{equation}
\begin{equation}
\label{conteq}
    f_\mathrm{cont} =\frac{ f_\mathrm{BB} -f_\mathrm{MB}(\Delta_\mathrm{MB}/\Delta_\mathrm{BB})}{1-\Delta_\mathrm{MB} / \Delta_\mathrm{BB}}
\end{equation}
for narrow-band and broad-band images in analogy to the previous work \citep[e.g.][]{koyama_panoramic_2010, tadaki_nature_2013, daikuhara_star-formation_2024}.
Line flux ($F_\mathrm{line}$) and continuum flux density ($f_\mathrm{cont}$) are described using the medium-band flux density. $\Delta$ denotes the FWHM value of each filter's response function. The error in each map is calculated by taking into account the propagation of errors in Equations \ref{lineeq} and \ref{conteq} using Root Mean Square (RMS) images from the F335M and F356W data.
We use empirical PSFs (released as part of \citealt{suess_medium_2024}) to match the PSF sizes between the medium-band and broad-band images for each filter combination.

We make radial profiles in the \oiii\ line map, the continuum map, and the F356W image using the \texttt{RadialProfile} class in the \texttt{Photutils} module with circler apertures. Additionally, we compute the effective radius and S\'{e}rsic index $n$ in those maps using \texttt{GALFIT} \citep{peng_detailed_2002, peng_detailed_2010}.

We obtain the line and continuum maps, shown in the left panels of Figure~\ref{fig:jwst_oiii}. 
The \oiii\ line map size appears comparable to the empirical PSF size, whereas the continuum map is observed to be more extended. Radial profiles are shown in the right panel of Figure~\ref{fig:jwst_oiii}.
The radial profiles of the continuum map and the F356W image are more extended than the PSF profile of the F356W image.
The comparison indicates that the stellar light is more extended than the PSF size, whereas the emission line map presents a more compact distribution. 

The line map is well characterized with the S\'{e}rsic profile of $r_e = 0.68^{+0.06}_{-0.06}$ kpc and $n = 0.55^{+0.16}_{-0.16}$, which is smaller than that for the continuum map ($r_e = 0.90^{+0.04}_{-0.04}$ kpc, $n = 0.70^{+0.08}_{-0.08}$).

From this comparison, we argue that the central $< 0.7$\,kpc region, is dominated by strong \hb+\oiii\ emissions, whereas the remaining part of the galaxy is dominated by older stellar populations, accounting for the observed Balmer break. We note that while the NIRSpec MSA slit (an open area of $0.20''\times0.46''$) is centered on the central region by design, it also covers the outer region, to $r\sim 1.4$\,kpc. The discrepancy between the line-based $A_V$ and the SED-based $A_V$ supports our view. 
The galaxy is likely in a phase that (has just) passed the primary star-forming phase, moving toward the quenching population with little residual star formation.

\subsubsection{Physical Interpretation of ``Outside-in" Quenching}
\label{Outside-in quenching}

The line-mapping analysis involving medium-band imaging for ID52153 shows that quenching is proceeding ``outside-in." Several previous works have shown evidence of outside-in quenching across a wide cosmological time.
In the local Universe, \citet{gallart_outside-disk_2008} presented an outside-in disk evolution in the Large Magellanic Cloud. \citet{pfeffer_age_2022} showed that galaxies in groups and clusters have similarly aged or younger inner regions using EAGLE simulation.
According to \citet{le_bail_jwstceers_2024}, one of the outside-in quenching scenarios is lopsidedness caused by major mergers. This is consistent with the post-star burst phase of the galaxy. 

Another scenario is a wet compaction event \citep[e.g.,][]{dekel_wet_2014,zolotov_compaction_2015}. During wet compaction, cold gas inflows efficiently into the center of the galaxy. This boosts the gas density and triggers high star formation rates in the core, culminating in a ``blue-nugget" phase \citep[e.g.,][]{lapiner_wet_2023}, since the center appears blue, dominated by young stars.
Meanwhile, the outer region experiences a decline in star formation. This phenomenon occurs because gas in the outskirts is either transported to the center or consumed there, thereby reducing the gas reservoir available for star formation in the outer region. 

By involving theoretical calculations, \citet{gelli_can_2024} discussed that SNe can suppress SF in low-mass galaxies. For the case of the $z = 7.3$ galaxy found in \citet{looser_recently_2024}, it is difficult to explain it only with the SNe feedback model. On the other hand, the galaxy at $z = 5.2$ found by \citet{strait_extremely_2023} can be explained by the SNe feedback model. The conclusion of that paper, quenching would be driven by not only SNe but also other different mechanisms, needed extra energy, such as outflow triggered by starbursts and AGN activities.

In the context of {\it massive} galaxy quenching, ``inside-out" quenching has been frequently discussed \citep[e.g.][]{genzel_sins_2014,tadaki_bulge-forming_2017,tacchella_dust_2018, Liu23}. This quenching hypothesizes that star formation quenching proceeds from the center of the galaxies first and then toward the outer region. Theoretical models including AGN feedback, morphological quenching and gas consumption, and depletion successfully reproduce this inside-out quenching pattern. The observed trend seen in ID52153 thus sheds new light on quenching in {\it low-mass}, high-redshift galaxies, which is a relatively unexplored area. While the existing data do not allow for the same line-map analysis for the other three galaxies, the observed spectroscopic features (i.e. Balmer break and dusty emission regions) are not inconsistent with the outside-in scenario.

The key question left is the nature of the central engine.
However, if AGN alone is the dominant source, one would expect a PSF-like profile, especially in the \oiii\ line map. We saw that the radial profile appears slightly elongated, suggesting that at least there exists an extended star-forming region. Ideally, IFU observations with a higher spectral resolution setup would discriminate different origins, further offering clues about past starburst activity too --- if younger stars (i.e. \ha) are confined to small, intense pockets, this would strengthen the ``bursty star-formation then fade" explanation.

\subsection{``Green flash" --- A key observable to low-mass galaxy quenching?}
\label{quenching_mechanism}

In what follows, we revisit a few possible physical mechanisms and discuss if those would reproduce the unique features observed in our samples --- all of our sample galaxies show a clear Balmer break, suggesting that their star formation ceased at least $\sim$100\,Myr ago, whereas they also show moderately strong emission lines. These two spectroscopic features characterize that these galaxies appear to be on the verge of quenching, whereas there exist {\it residual} line emissions somewhere in the same system. The coexistence of the two features is actually seen at lower redshifts in relatively evolved but not completely quenched galaxies, i.e. post-starburst \citep[e.g.,][]{dressler_rotational_1983,yesuf_starburst_2014}. Galaxies are known to spend a relatively short amount of time in such a phase, also known as the green valley, transitioning into quiescence. Being in a similar phase, our sample galaxies are ideal to speculate the key physical mechanisms responsible for early galaxy quenching in the low-mass regime.  

Firstly, powerful outflows can reduce or even expel the interstellar medium (ISM) of a galaxy out of the potential well, leading to star formation quenching. Such outflows can be driven by intense star formation or AGN activity. In low-mass galaxy systems, the relative impact from such outflows on the galaxy scale could be significant, due to the shallow potential well; conversely, in massive galaxies, the impact from outflows is expected to be relatively small.

A typical indication of outflows from quiescent galaxies is the presence of broad or offset emission-line components, as observed in ground-based studies \citep[e.g., Keck/MOSFIRE and Subaru/MOIRCS,][]{kubo_agn_2022} and recent JWST observations \citep[e.g.,][]{belli_star_2024, de_graaff_efficient_2024}.
Our spectra do not have sufficient resolution to reveal clear velocity shifts, making it difficult to investigate the presence of ongoing outflows for these four galaxies. Exceptionally, ID52153 exhibits a compact emission-line region (Sec.~\ref{sec:linemap}). Additionally, the younger stellar population appears to be more concentrated in the inner region, whereas older stars inhabit the outskirts.

Second, radiative heating from massive stars in the HII regions, supernova shocks, and AGNs can suppress star formation \citep[e.g.,][]{hopkins_galaxies_2014,hopkins_radiative_2020,xie_radiative_2017}. 
When the interstellar medium is heated to the order of a few thousand Kelvins, dust grains start to be destroyed. The gas therein then becomes ionized or thermally unstable. As a result, star formation is suppressed because the cool, dense pockets of gas necessary for stellar birth are disrupted \citep{krumholz_which_2011}.

Dust grains typically sublimate at temperatures above 2,000\,K. If our sample galaxies host intense radiation fields (either from short-lived starbursts or an AGN), this may explain why they appear dustier in emissions while also showing the sign of star formation quenching in the stellar continuum break. 
Dust in the central region can provide effective self-shielding, allowing star formation to continue. In contrast, in the outer regions, heating by low-mass stars could inhibit the formation of molecular hydrogen, thereby suppressing star formation \citep[e.g.,][]{kajisawa_dust_2015}. As such, the presence of a compact dusty core, as seen in ID52153, may play a key role in quenching while showing the unique spectroscopic features observed.

The third mechanism involves rapid gas consumption through intense, short-lived bursts of star formation. In this burst-and-fade picture, a galaxy forms stars at a very high rate for a relatively brief period, quickly using up much of its available cold gas \citep{wild_star_2020}. 
Observations of dusty star-forming galaxies, especially at high redshifts, show that they often experience elevated star formation above the main sequence \citep[e.g.,][]{casey_dusty_2014,madau_cosmic_2014}. Given that our galaxies are located on and below the star formation main sequence (Figure\,\ref{sfms}), the observed moderate emissions may well be the residual star formation fading from such a bursty phase. 

An open question is whether the gas is completely consumed or not. For example, gas blown away by strong outflow may be retained later and may trigger subsequent star formation. In fact, several studies reported a reservoir of molecular gas within massive quiescent galaxies at $z>1$ \citep{gobat_unexpectedly_2018,whitaker_quenching_2021,magdis_interstellar_2021,morishita_compact_2022}.
Submillimeter facilities, such as ALMA, would be vital for measuring molecular gas reservoirs and investigating the rapid gas consumption scenario. 

Lastly, environments can influence star formation in low-mass galaxies. \citet{morishita_accelerated_2025} reported the discovery of an overdensity at $z\sim5.7$ in the same field where our samples were drawn. Among our samples, ID53079 ($z = 5.78$) and ID58043 ($z = 5.67$) were identified as members of this overdensity. Because they reside in an overdense region, one possible scenario is that they underwent a burst of star formation, which then ceased in a relatively short amount of time. In contrast, ID52153 ($z = 5.54$) and ID85007 ($z = 5.10$) do not appear to be part of the discovered overdensity. As such, we conclude that the environment alone is insufficient to explain the observed properties. 

We note that our samples in this study were selected on the basis of the availability of JWST/NIRSpec PRISM data. However, it remains to be determined whether the differences in dust attenuation between emission-line regions and the overall stellar component are common to other Balmer-break galaxies at $z\sim5$ that exhibit emission lines. We anticipate that enlarging the sample in future studies will clarify this question.

\section{Conclusions}
\label{sec:conclusions}
We identified four Balmer-break galaxies at $z\sim5$--6 from recent JWST spectroscopic surveys in the Abell~2744 field. Uniquely, all of these Balmer-break galaxies show strong \ha\ emission but weaker \hb\ than predicted from the Case-B recombination model. 

From our line fitting analysis and SED analysis involving the entire galaxy system, we conclude that these line emitting regions are more attenuated compared to the overall stellar component of the host, indicating differential dust extinction.. 

An in-depth analysis of one of the sample galaxies, ID52153, using medium- and broad-band imaging revealed that the \oiii\ emission is compact and centrally concentrated (effective radius $\sim0.7$ kpc), while the stellar continuum responsible for the Balmer break is more extended (effective radius $\sim0.9$ kpc). The spatial decoupling of the compact, young/active star-forming central region and the extended, older/quenched outer region in ID52153 suggests an ``outside-in" quenching scenario, where star formation ceased in the outskirts before the central region. This contrasts with the ``inside-out" quenching typically seen in massive galaxies, and could be driven by processes like wet compaction or mergers. While the exact nature of the central activity (star formation vs. AGN) remains uncertain, the evidence points towards that these high-redshift, low-mass galaxies quench via a pathway that preserves residual activity in a compact, dusty core.

Several quenching mechanisms responsible for the observed properties were discussed: powerful outflows (difficult to confirm with current resolution), radiative heating (potentially explaining dusty emission coexisting with quenching), rapid gas consumption after a burst (consistent with being below the main sequence), and environmental influence (relevant for two galaxies in an overdensity, but not all).

Given that our study focuses on a small number of galaxies at a relatively narrow redshift range, further investigations with a larger sample would be essential for a comprehensive understanding of the quenching mechanism in act in low-mass galaxies in the early Universe.

\begin{acknowledgments}

We thank GLASS-JWST team, UNCOVER team, GO3073 \citep{castellano_jwst_2024} program and Dr. Akos Bogdan for making their data available. We thank Dr. Roberts-Borsani for kindly making their reduced spectra available for our study. 
The spectrum of ID85007 is obtained in the program JWST-GO-3073. This is based on observations made with the NASA/ESA/CSA JWST. The JWST data presented in this article were obtained from the Mikulski Archive for Space Telescopes (MAST) at the Space Telescope Science Institute. The specific observations analyzed are associated with program JWST-GO-3073 and can be accessed via \dataset[DOI]{https://doi.org/10.17909/4r6b-bx96}.
We thank Dr. Mariko Kubo for the valuable discussion. 
KT and KD are supported by KAKENHI (International Leading Research) and Graduate Program on Physics for the Universe (GP-PU), Tohoku University. 
TM received support from NASA through the STScI grants HST-GO-17231, JWST-GO-1747, and JWST-GO-3990.
TK acknowledges financial support from JSPS KAKENHI Grant Numbers 24H00002 (Specially Promoted Research by T. Kodama et al.) and 22K21349 (International Leading Research by S. Miyazaki et al.).
This work was also supported by JSPS Core-to-Core Program (grant number: JPJSCCA20210003). 
ZL acknowledges support from JSPS KAKENHI Grant Number 24KJ0394.
\end{acknowledgments}

\vspace{5mm}


\software{
\texttt{Numpy} \citep{harris_array_2020},
\texttt{Astropy} \citep{the_astropy_collaboration_astropy_2013},
\texttt{Bagpipes} \citep{carnall_inferring_2018, carnall_vandels_2019},
\texttt{CIGALE} \citep{boquien_cigale_2019},
\texttt{emcee} \citep{foreman-mackey_emcee_2013}, 
\texttt{GALFIT} \citep{peng_detailed_2002, peng_detailed_2010},
\texttt{CIAO} \citep{fruscione_ciao_2006}
}

\begin{table*}
	\centering
	\caption{Properties of sample Balmer break galaxies.}
	\label{value}
	\begin{tabular}{lcccccccc} 
		\hline
        \hline
  ID&R.A.&Dec.&$z_\mathrm{spec}$&$\mu$&$\log_{10}(M_{\ast})$&$\log_{10}(\mathrm{SFR_{10Myr}})$&$A_V$&mass weighted age\\
  &&&&&$\mathrm{M}_{\odot}$&$\mathrm{M}_{\odot}/\mathrm{yr}$&[mag]&[Gyr]\\
		\hline
  52153&3.57006&-30.40369&$5.546$&$2.1^{+0.1}_{-0.1}$&$8.79^{+0.06}_{-0.05}$&$0.72^{+0.13}_{-0.13}$&$0.61^{+0.06}_{-0.06}$&$0.12^{+0.09}_{-0.05}$\\
  53709&3.59614&-30.37138&$5.786$&$2.3^{+0.1}_{-0.1}$&$9.15^{+0.03}_{-0.02}$&$0.18^{+0.11}_{-0.11}$&$0.35^{+0.05}_{-0.04}$&$0.15^{+0.05}_{-0.02}$\\
  58043&3.59316&-30.34647&$5.663$&$2.0^{+0.1}_{-0.1}$&$9.22^{+0.01}_{-0.01}$&$0.05^{+0.16}_{-0.23}$&$0.13^{+0.03}_{-0.02}$&$0.07^{+0.01}_{-0.00}$\\
  85007&3.46853&-30.32615&$5.106$&$-$(1.0)&$8.84^{+0.03}_{-0.03}$&$0.04^{+0.13}_{-0.15}$&$0.14^{+0.06}_{-0.06}$&$0.23^{+0.04}_{-0.06}$\\
  \hline
  \hline
	\end{tabular}
\end{table*}

\begin{table*}
\label{line_fit}
	\centering
	\caption{The results of emission line fittings. These fluxes are observed fluxes. Error values are 1-$\sigma$ errors.}
	\label{linefittingvalue}
	\begin{tabular}{lcccc} 
		\hline
        \hline
ID&$F_\mathrm{H\beta}$&$F_\mathrm{[O~III]\lambda5007}$&$F_\mathrm{H\alpha}$&$F_\mathrm{[N~II]\lambda6583}$\\

&$[\mathrm{erg/s/cm^2}]$&$[\mathrm{erg/s/cm^2}]$&$[\mathrm{erg/s/cm^2}]$&$[\mathrm{erg/s/cm^2}]$\\
		\hline
52153&$<3.00 \times 10^{-19}$ (2$\sigma$)&$2.11^{+0.20}_{-0.19} \times 10^{-18}$&$>5.31 \times 10^{-18}$ (2$\sigma$)&$<8.28 \times 10^{-19}$ (2$\sigma$)\\
  53709 &$3.95^{+4.17}_{-2.28} \times 10^{-19}$&$4.43^{+0.12}_{-0.12} \times 10^{-18}$&$>5.28 \times 10^{-18}$ (2$\sigma$)&$<8.74 \times 10^{-19}$ (2$\sigma$)\\
  58043 &$<3.14 \times 10^{-19}$ (2$\sigma$)&$1.29^{+0.28}_{-0.27} \times 10^{-18}$&$>1.65 \times 10^{-18}$ (2$\sigma$)&$<1.86 \times 10^{-18}$ (2$\sigma$)\\
  85007 &$<4.10 \times 10^{-19}$ (2$\sigma$)&$6.61^{+0.33}_{-0.33} \times 10^{-18}$&$>4.98 \times 10^{-18}$ (2$\sigma$)&$<4.92 \times 10^{-18}$ (2$\sigma$)\\
  \hline
        \hline
	\end{tabular}
\end{table*}

\bibliography{mylib,newbib}
\bibliographystyle{aasjournal}
\end{document}